\documentstyle[12pt]{article}

\begin{document}

\begin{center}
{\large {\bf ACOUSTIC WORMHOLES}}\\[0pt]

\vspace{8mm}

Kamal Kanti Nandi,$^{a,b,}$\footnote{%
E-mail: kamalnandi@hotmail.com} Yuan-Zhong Zhang$^{b,}$\footnote{%
E-mail: yzhang@itp.ac.cn}\\[0pt]
and Rong-Gen Cai$^{c,b,}$\footnote{%
E-mail: cairg@itp.ac.cn}

\vspace{5mm} {\footnotesize {\it \
 $^a$ Department of Mathematics,
University of North Bengal, Darjeeling (W.B.) 734 430, India\\[0pt]
$^b$Institute of Theoretical Physics, Chinese Academy of Sciences,
P.O.Box 2735, Beijing 100080, China\\[0pt]
$^c$ CASPER, Department of Physics, Baylor University, Waco,
TX76798-7316, USA }}
\end{center}

\begin{abstract}
Acoustic analogs of static, spherically symmetric massive
traversable Lorentzian wormholes are constructed as a {\em formal}
extension of acoustic black holes. The method is straightforward
but the idea is interesting in itself. The analysis leads to a new
acoustic invariant for the massless counterpart of the
Einstein-Rosen model of an elementary particle. It is shown that
there is a marked, in a sense even counterintuitive, physical
difference between the acoustic analogs of black holes and
wormholes. The analogy allows us to also portray the nature of
curvature singularity in the acoustic language. It is demonstrated
that the light ray trajectories in an optical medium are the same
as the sound trajectories in its acoustic analog. The implications
of these analogies in the laboratory set up and in the different
context of phantom energy accretion have been speculated.

\bigskip

\noindent PACS number(s): 04.20.Gz, 04.50.+h, 04.70.Dy, 47.90.+a
\end{abstract}

\vspace{8mm}

\section{Introduction}

Several years ago, Unruh [1] proposed a novel idea as to how a sonic horizon
in transonic flow could give out a thermal spectrum of sound waves mimicking
Hawking's general relativistic black hole evaporation. The radiation of such
sound waves is now commonly known as Hawking-Unruh radiation [2]. This basic
acoustic analogy has been further explored in later years and under
different physical circumstances by several authors [3-11]. A very useful
account and a detailed extension of some of these developments can be found
in Visser [4] wherein it is also demonstrated that the Hawking radiation is
purely a kinematic effect quite independent of the machinery of Einstein's
field equations. This information is consistent with the fact that a
detector with uniform acceleration $a$ in vacuum responds as though it were
immersed in a thermal bath of temperature $T=\hbar a/2\pi \kappa c$, where $%
\kappa $ is Boltzmann's constant, $\hbar $ is Planck's constant and $c$ is
the speed of light in vacuum [2]. This is the same as Hawking temperature
when one formally puts $a\equiv g$, where $g$ stands for the gravitational
acceleration at the black hole surface. All in all, the past and present
works surrounding the basic analogy have engendered the practical
possibility of detecting Hawking-Unruh radiation in fluid (especially
superfluid) models under appropriately simulated conditions [8-11]. Quite
reasonably, there is a widespread current interest in this topic as these
models open up an alternative window to look for many unknown effects of
quantum gravity on black holes in the laboratory.

As things stand, the theoretical acoustic modelling can be achieved in two
reciprocal ways: The first way is to start from the fluid equations for
continuity, Euler equations and the equation of state. Linearize them and
obtain the wave equation for sound waves. This process allows us to
equivalently describe sound waves as propagating on a spacetime metric,
called the acoustic metric. The equation of propagation can be cast into
that of a massless minimally coupled scalar field [1,4]. Following this
method, Visser [4] obtained the Painlev\'{e}-Gullstrand form of the
Schwarzschild exterior metric up to a conformal factor and also the
canonical acoustic metric. In that work, various types of acoustic regions
are consistently defined leading to an improved formula for Hawking-Unruh
radiation. The second way, though it is not the one usually followed, is to
start from a given spacetime metric of general relativity and obtain an
equivalent fluid description in terms of density, pressure, equation of
state etc. This new method has been worked out in a recent paper by Visser
and Weinfurtner [6] in the context of building the acoustical analog of the
equatorial slice of a Kerr black hole. The analogy enables one to gain
insight into Kerr spacetime itself as well as into the vortex-inspired
experiments. All the above works have led to a systematic development of
what might be generally called the physics of optical or acoustical black
holes.

In this paper, a first hand idea of acoustic wormholes is advanced as a
natural extension of acoustic black holes although it is known that the
intrinsic topologies of black and wormholes are widely different. After
briefly re-deriving the standard form of acoustic metric in Sec.2, we
present, in Sec.3, a class of static, spherically symmetric, traversable
Lorentzian wormholes of general relativity. In Sec.4, we adopt the second
method to develop the acoustic analog of these wormholes including the zero
mass limit and compare their characters with those of the acoustic analog of
the Schwarzschild black hole. Geodesic equations are discussed in Sec.5. A
summary together with some speculative remarks appear in Sec.6.

\section{Acoustic metric from fluid equations}

For completeness, let us first survey in brief how the acoustic metric is
developed. Consider an irrotational, nonrelativistic, inviscid, barotropic
fluid so that the equations are

\begin{equation}
\overrightarrow{\nabla }\times \overrightarrow{v}=0\Rightarrow
\overrightarrow{v}=\overrightarrow{\nabla }\Psi
\end{equation}

\begin{equation}
\frac{\partial \rho }{\partial t}+\overrightarrow{\nabla }.\left( \rho
\overrightarrow{v}\right) =0
\end{equation}%
\begin{equation}
\rho \left[ \frac{\partial \overrightarrow{v}}{\partial t}+(\overrightarrow{v%
}.\overrightarrow{\nabla })\overrightarrow{v}\right] =-\overrightarrow{%
\nabla }p
\end{equation}

\begin{equation}
p=p(\rho )
\end{equation}%
in which all relevant terms have their usual meanings. Now linearize the
equations around a background exact solution set [ $p_{0}(t,\overrightarrow{x%
}),\rho _{0}(t,\overrightarrow{x}),\Psi _{0}(t,\overrightarrow{x})$] such
that%
\begin{equation}
p=p_{0}+\delta p_{0}+0(\delta p_{0})^{2},\rho =\rho _{0}+\delta \rho
_{0}+0(\delta \rho _{0})^{2},\Psi =\Psi _{0}+\delta \Psi _{0}+0(\delta \Psi
_{0})^{2}
\end{equation}%
where \ $\delta $ denotes a small perturbation to the relevant quantities.
Then the perturbations satisfy the equation [1]

\begin{equation}
-\partial _{t}\left[ \frac{\partial \rho }{\partial p}\rho _{0}\left(
\partial _{t}\Psi _{1}+\overrightarrow{v}_{0}.\overrightarrow{\nabla }\Psi
_{1}\right) \right] +\overrightarrow{\nabla }.\left[ \rho _{0}%
\overrightarrow{\nabla }\Psi _{1}-\frac{\partial \rho }{\partial p}\rho _{0}%
\overrightarrow{v}_{0}\left( \partial _{t}\Psi _{1}+\overrightarrow{v}_{0}.%
\overrightarrow{\nabla }\Psi _{1}\right) \right] =0.
\end{equation}%
This equation can be neatly rewritten as the minimally coupled wave equation
[4]

\begin{equation}
\frac{1}{\sqrt{-\overline{g}}}\frac{\partial }{\partial x^{\mu }}\left(
\sqrt{-\overline{g}}\overline{g}^{\mu \nu }\frac{\partial \Psi _{1}}{%
\partial x^{\nu }}\right) =0
\end{equation}%
where $\overline{g}_{\mu \nu }$ is the acoustic metric, $\Psi _{1}\equiv
\delta \Psi _{0}$, $\overline{g}=\det \left\vert \overline{g}_{\mu \nu
}\right\vert $, and

\begin{equation}
ds_{acoustic}^{2}=\overline{g}_{\mu \nu }dx^{\mu }dx^{\nu }=\frac{\rho _{0}}{%
c_{s}}\left[ -c_{s}^{2}dt^{2}+\left( dx^{i}-v_{0}^{i}dt\right) \delta
_{ij}\left( dx^{j}-v_{0}^{j}dt\right) \right]
\end{equation}%
in which $i,j=1,2,3$ and the speed of sound is given by $c_{s}^{2}=\frac{%
dp_{0}}{d\rho _{0}}$. This is geometrization of acoustics. Our strategy in
this paper is to follow the reverse route, that is, we cast a given metric
of general relativity into the form (8) and find out the fluid variables for
an analogous acoustic configuration. Before closing this section, we note
that in a physical situation where $\overrightarrow{v}_{0}=0$, and $\rho
_{0}/c_{s}$= constant, we have an exact Minkowski spacetime where the role
of signal speed is played by that of sound waves. This phenomenon calls for
some sort of sonic special relativity where the observers are equipped with
only sound waves, that is, a world where the observers can only
\textquotedblleft hear\textquotedblright\ but not \textquotedblleft
see\textquotedblright . This scenario has been explicitly recognized and
contemplated upon by Visser [4]. It is of some historical interest to note
that such a sonic relativity was in fact conceived decades ago where the
subsonic and supersonic Lorentz-like transformations contained $c_{s}$ as
the invariant speed [12]. At that time, it was advanced not as a formal
theory but rather as a technical tool for solving various acoustical
problems in linearized sub- and supersonic aerodynamics including the
Earth's magnetospheric bow shock geometry. Recently, these concepts have
received additional support, though not yet a confirmation, in the current
context of classical acoustical black holes that fundamentally issue forth
from fluid equations {\it ab initio}. However, it is quite likely that
quantum effects could destroy the inherent Lorentz invariance at the
microscopic level [13].

\section{Wormhole solutions of general relativity}

Let us consider the field equations of the Einstein minimally coupled scalar
field theory which are given by (We adopt units $8\pi G=c=\hbar =1$, and
signature convention $-,+,+,+$ ):%
\begin{equation}
R_{\mu \nu }-\frac{1}{2}g_{\mu \nu }R=T_{\mu \nu }
\end{equation}

\begin{equation}
T_{\mu \nu }=\alpha \left[ \Phi _{,\mu }\Phi _{,\nu }-\frac{1}{2}g_{\mu \nu
}\Phi _{,\sigma }\Phi ^{,\sigma }\right]
\end{equation}

\begin{equation}
\Phi _{;\mu }^{;\mu }=0
\end{equation}%
where $\alpha $ is an arbitrary constant that can have any sign, $\Phi $ is
the scalar field, $R_{\mu \nu }$ is the Ricci tensor and the semicolon
denotes covariant derivatives with respect to the metric $g_{\mu \nu }$. If $%
\alpha $ is negative, or $\Phi $ is imaginary, then the stress tensor $%
T_{\mu \nu }$ has an overall negative sign so that it represents exotic
matter necessary for threading wormholes [14]. We consider a solution of the
form

\begin{equation}
ds^{2}=g_{\mu \nu }dx^{\mu }dx^{\nu }=-e^{-2\phi (r)}dt^{2}+e^{-2\psi (r)}
\left[ dr^{2}+r^{2}d\theta ^{2}+r^{2}\sin ^{2}\theta d\varphi ^{2}\right]
\end{equation}

\begin{equation}
\phi (r)=-\beta \ln \left[ \frac{1-\frac{m}{2r}}{1+\frac{m}{2r}}\right]
,\psi (r)=(\beta -1)\ln \left( 1-\frac{m}{2r}\right) -(\beta +1)\ln \left( 1+%
\frac{m}{2r}\right)
\end{equation}

\begin{equation}
\Phi (r)=\left[ \frac{2(1-\beta ^{2})}{\alpha }\right] ^{\frac{1}{2}}\ln %
\left[ \frac{1-\frac{m}{2r}}{1+\frac{m}{2r}}\right]
\end{equation}%
where the two undetermined constants $m$ and $\beta $ are related to the
source strengths of the gravitational and scalar parts of the configuration.
The solution was first put in that form by Buchdahl [15] and it describes
all weak field tests of gravity just as accurately as does the Schwarzschild
metric. The conserved mass of the configuration is given by%
\begin{equation}
M=m\beta .
\end{equation}%
Once the scalar component is set to zero ( $\beta =1\Rightarrow \Phi =0$),
the solutions (13), (14) reduce to the Schwarzschild black hole in
accordance with Wheeler's \textquotedblleft no scalar
hair\textquotedblright\ conjecture. Physically, this indicates the
possibility that the scalar field be radiated away during collapse so that
the end result is a Schwarzschild black hole [16]. The class of solutions
(13), (14) describes asymptotically flat traversable wormholes with finite
tidal forces [17]. The throat appears at the coordinate radii%
\begin{equation}
r_{0}^{\pm }=\frac{m}{2}\left[ \beta {\pm }\sqrt{\beta ^{2}-1}\right] ,
\end{equation}%
that connects two asymptotically flat regions, one at $r=\infty $ and the
other appearing at $r=0.$ Here we take only the positive sign ($r_{0}^{+}$).
To find the second asymptotic region, note that, under the transformation $r=%
\frac{m^{2}}{4\xi }$, the metric $ds^{2}(r)\rightarrow (-1)^{2\beta
}ds^{2}(\xi )$. Thus, in order to ensure the form invariance of the metric
including the signature, we must have $\beta \in N$. It is obvious that the
metric becomes flat also at $\xi =\infty $, that is, at $r=0$. The $r-$ and
the $\xi -$coordinates meet at $r=\xi =m/2.$ In other words, either $r$ or $%
\xi $ can be taken as a global coordinate patch. The requirement that the
throat radii be real implies that $\beta ^{2}>1$. Under this wormhole
condition, consider the two mutually exclusive possibilities for $\Phi $: It
could be either real or imaginary. The reality of \ $\Phi $ [more precisely,
the reality of the scalar charge $M_{S}=-m\left[ 2(1-\beta ^{2})/\alpha %
\right] ^{\frac{1}{2}}$obtained from $\Phi \simeq M_{S}/r$] demands that $%
\alpha <0$. Alternatively, one could have $\alpha >0$ and allow for an
imaginary $\Phi $. The latter choice presents no pathology or inconsistency
in the wormhole physics, as is recently shown in Ref.[18]. In both cases,
however, we have a negative sign before the stress tensor (10) and
consequently almost all energy conditions are violated. In what follows, we
need to concentrate on the metric part only. Under this setting, we now
proceed to find the acoustic model of the wormhole.

\section{Acoustic wormholes}

With fluid, it is {\it not} possible to set up the geometry having the {\it %
topological appearance }of a wormhole that is opening at two ends to
asymptotically flat spacetimes. What one can do is to examine how the fluid
quantities behave at the corresponding coordinate values ($r=\infty
,r=r_{0}^{+},r=0$) in the wormhole geometry. The situation is quite similar
to the acoustic analogue of Schwarzschild vacuum which is modelled by real
(even flowing) fluid filling all space. It is not possible to find also the
topology (e.g., the double cone in the Kruskal-Szekeres extension) in the
fluid field. With this limitation in mind, consider the acoustic metric (8)
and choose $\overrightarrow{v}_{0}=0$. Then formally replace the vacuum
speed of light $c$ in (12) [which we have set to unity] by the asymptotic
speed of sound $c_{\infty }$ corresponding to a linear medium, equate the
metric (12) with the metric (8), that is, set $\overline{g}_{\mu \nu
}=g_{\mu \nu }.$ Immediately one obtains the density and sound speed
profiles respectively as

\begin{equation}
\rho _{0}(r)=\rho _{\infty }\left( 1-\frac{m^{2}}{4r^{2}}\right)
\end{equation}

\begin{equation}
c_{s}(r)=c_{\infty }\left( \frac{1-\frac{m}{2r}}{1+\frac{m}{2r}}\right)
^{2\beta }\times \left( 1-\frac{m^{2}}{4r^{2}}\right) ^{-1}.
\end{equation}%
It must be emphasized again that we are looking for an acoustical analog of
the wormhole. Consequently, the fluid density $\rho _{0}$ is {\it not} the
exotic energy density of the scalar field $\Phi $ which is actually negative
for $\beta ^{2}>1$ and is given by

\begin{equation}
\rho (\Phi )=-\left[ \frac{256m^{2}r^{4}(\beta ^{2}-1)(1-\frac{m}{2r}%
)^{2\beta }(1+\frac{m}{2r})^{-2\beta }}{(m^{2}-4r^{2})^{4}}\right] .
\end{equation}%
The same statement applies also to pressure. Eqs.(7) and (11) imply that,
formally, $\Psi _{1}\equiv \Phi $ but the underlying physics is entirely
different: $\Psi _{1}$ represents a fluid perturbation propagating with
local speed $c_{s}$ on top of a spacetime $\overline{g}_{\mu \nu }$ while
the dilaton $\Phi $ mediates a medium range force similar to the
electromagnetic force that propagates with the speed of light $c$. For a
more elaborate discussion on this point, see Ref.[19]. Returning to the
fluid description, we can rewrite the Euler equation (3)
for $p_{0}=p_{0}(r)$
by explicitly displaying the force required to hold the fluid configuration
in place against the pressure gradient:

\begin{equation}
\overrightarrow{f}=\rho _{0}\left(
\overrightarrow{v}_{0}.\overrightarrow{ \nabla }\right)
\overrightarrow{v}_{0}+c_{s}^{2}\partial _{r}\rho _{0}
\widehat{r}.
\end{equation}

It is now possible to find the pressure profile by integrating the
 equation $(\overrightarrow{v}_{0}=0)$:

\begin{equation}
f_{\widehat{r}}=\frac{dp_{0}}{dr}=c_{s}^{2}\frac{d\rho _{0}}{dr}
\end{equation}%
which gives, under the physical condition that $p_{\infty }\rightarrow 0$,

\begin{equation}
p_{0}(r)=8\rho _{\infty }c_{\infty }^{2}\left[
\frac{-1+(2r-m)^{4\beta -1}(2r+m)^{-(4\beta
+1)}(m^{2}+4r^{2}+16mr\beta )}{16(16\beta ^{2}-1)}\right]
\end{equation}

The barotropic equation of state can be found by eliminating $r$ from
Eqs.(17) and (22) as

\begin{equation}
p_{0}(\rho _{0})=\frac{\rho _{\infty }c_{\infty }^{2}}{2(16\beta ^{2}-1)}%
\left[ -1+\left( \frac{1-\Omega }{\Omega }\right) ^{4\beta -1}\left( \frac{%
1+\Omega }{\Omega }\right) ^{-(4\beta +1)}\left( 1+\frac{8\beta }{\Omega }+%
\frac{1}{\Omega ^{2}}\right) \right]
\end{equation}%
where $\Omega \equiv \left( 1-\frac{\rho _{0}}{\rho _{\infty }}\right) ^{%
\frac{1}{2}}.$ The corresponding results for the Schwarzschild black hole
can be obtained by setting $\beta =1$ for which there is a horizon at $%
r=r_{H}=m/2$. It is of interest to note that $\rho _{0}(r)$ does not involve
the scalar field parameter $\beta $ at all, and it is of the same form as in
the Schwarzschild case [6]. At the throat of the wormhole, the pressure is
given by

\begin{equation}
p_{0}(r_{0}^{+})=\frac{\rho _{\infty }c_{\infty }^{2}}{2(16\beta ^{2}-1)}%
\left[ -1+10\beta \Pi \left( -1+\Pi \right) ^{4\beta -1}\left( 1+\Pi \right)
^{-(4\beta +1)}\right]
\end{equation}%
where $\Pi \equiv \beta +\sqrt{\beta ^{2}-1}.$ The equations (17), (18),
(22)-(24) represent the exact acoustic model for the wormhole for $\beta >1$%
 From the expression (22), it follows that the pressure drops from zero at
infinity to negative values as one proceeds to the throat. This phenomenon
is similar to that occurring in acoustic black holes. By varying $\beta $,
the negative value can be brought as near to zero as we please. For
instance, as $\beta \rightarrow \infty $, we have $p_{0}(r_{0}^{+})%
\rightarrow 0-$ . This implies that the fluid pressure increases to zero as
we enlarge the throat ($r_{0}^{+}$) of the wormhole. The limit $\beta
\rightarrow \infty $ also implies that the mass of the wormhole $M$ \ be
large. This property of pressure enhancement proportional to the enlargement
of mass is a generic property (it actually occurs for any value of $\beta >1$%
) of the class of wormholes under consideration here. However, with regard
to the overall energy condition, normalizing $\rho _{\infty }=c_{\infty }=1$%
, we can see that $\rho _{0}(r)+p_{0}(r)>0$ for $r\in \lbrack
r_{0}^{+},\infty )$ and $\rho _{0}(r)+p_{0}(r)\rightarrow 1$ rapidly as $%
r\rightarrow \infty .$ Also, $\rho _{0}(r_{0}^{+})+p_{0}(r_{0}^{+})>0$ for $%
\beta >1$ and $\rho _{0}(r_{0}^{+})+p_{0}(r_{0}^{+})\rightarrow 1$ very fast
as $\beta \rightarrow \infty $, that is, there is no violation of energy
condition either at the throat or elsewhere whatever be the size of the
wormhole. In contrast, for the acoustic analog of the Schwarzschild black
hole, $\rho _{0}(r_{H})+p_{0}(r_{H})<0$ implying that the null energy
condition is violated at the horizon. Roughly at $r=0.51m,$ $\rho
_{0}+p_{0}=0$ and for $r>0.51m$, we have $\rho _{0}+p_{0}>0$. Thus, there is
an extremely thin layer of exotic acoustic material of the order of
thickness $0.01m$ around the horizon. This energy violating behavior of the
acoustic analogue of Schwarzschild black holes is very different from that
of traversable wormholes. A separate detailed examination of other variants
of energy conditions would be worthwhile. We had, in the above assumed, for
simplicity, $\overrightarrow{v}_{0}=0$, but we could as well introduce any
velocity profile $v_{0}=v_{0}(r)$ in which case only the pressure profile $%
p_{0}=p_{0}(r)$ would be different from and perhaps more complicated than
Eq.(22).

We tabulate for a better view the profiles of fluid quantities both for
wormhole and black hole (with $\rho _{\infty }=c_{\infty }=1$):

\begin{center}
$%
\begin{array}{cccc}
\beta >1: & r=\infty & r=r_{th} & r=0 \\
\rho _{0}: & 1 & \rho _{0}(r_{0}^{+}) & \infty \\
c_{s}: & 1 & c_{s}(r_{0}^{+}) & 0 \\
p_{0}: & 0 & p_{0}(r_{0}^{+}) & -\frac{1}{16\beta ^{2}-1}%
\end{array}%
\begin{array}{cccc}
\beta =1: & r=\infty & r=r_{H} & r=0 \\
& 1 & 0 & \infty \\
& 1 & 0 & 0 \\
& 0 & -\frac{1}{30} & 0%
\end{array}%
$
\end{center}

It is only the overall profile, as evidenced above, that distinguishes
between black and wormholes. Of particular interest is the zero mass case ($%
M=m\beta =0$). This case actually illustrates the massless counterpart of
what was originally conceived by Einstein and Rosen in their two-sheeted
elementary particle model [20] with the so called \textquotedblleft
bridge\textquotedblright\ (throat) representing the positive mass particle.
This original concept has later been extended by Wheeler [21] and discussed
well enough in the literature [18, 21-23]. To arrive at this massless case,
we can set $\beta =0$, $m\neq 0$ in Eqs.(17),(18) and (22). (The other
alternative, $\beta \neq 0$, $m=0$ is trivial). We choose $m=-im%
{\acute{}}%
$ where $m%
{\acute{}}%
>0$, so that $r_{0}^{+}=m%
{\acute{}}%
/2$ . Then the relevant fluid profiles are

\begin{equation}
\rho _{0}(r)=\rho _{\infty }\left( 1+\frac{m%
{\acute{}}%
^{2}}{4r^{2}}\right) ,c_{s}(r)=c_{\infty }\left( 1+\frac{m%
{\acute{}}%
^{2}}{4r^{2}}\right) ^{-1},p_{0}(r)=\rho _{\infty }c_{\infty }^{2}\left( 1+%
\frac{4r^{2}}{m%
{\acute{}}%
^{2}}\right) ^{-1}.
\end{equation}%
We immediately find that

\begin{equation}
\rho _{0}c_{s}=\rho _{\infty }c_{\infty }=cons\tan t
\end{equation}%
which illustrates a {\it new} invariant of the particle model in its
acoustical analog. Such an invariant is unavailable in the massive case.
Moreover, at $r=r_{0}^{+}=m%
{\acute{}}%
/2$ , we can see that $p_{0}(r_{0}^{+})=+\rho _{\infty }c_{\infty }^{2}/2$, $%
\rho _{0}(r_{0}^{+})=2\rho _{\infty }$. Actually, there is a sharp {\it %
increase} in pressure at the throat from the asymptotic zero value implying
a condensation of fluid at the \textquotedblleft bridge\textquotedblright\
which seems to tally well with the particle model. This feature is in
contradistinction to the pressure {\it decrease} from the asymptotic zero
value to any negative value at the throat or to $p_{0}(r_{H})=-\rho _{\infty
}c_{\infty }^{2}/30$ at the horizon ($r_{H}=m/2$) in the acoustic analogs of
a massive wormhole or Schwarzschild black hole respectively. Also, $\rho
_{0}(r)+p_{0}(r)>0$ for $r\in \lbrack m%
{\acute{}}%
/2,\infty )$ showing that it is possible to model the Einstein-Rosen
particle with ordinary, as opposed to exotic, matter satisfying a simple
equation of state $p_{0}(\rho _{0})=\rho _{\infty }c_{\infty }^{2}(1-\rho
_{\infty }/\rho _{0}).$

As an aside, it is of interest to see what the acoustic analog looks like at
the naked singularity that occurs at $r_{NS}=m/2$ for $\beta <1$. (Note
that, for $\beta >1$, one has from Eq.(19), $\rho (\Phi )<0$, and the
geometry inevitably is that of a wormhole so that the minimum distance to
the center that the travellers can get to is $r_{0}^{+}>r_{NS}$. That is,
the singular surface is practically inaccessible to them.) We see from the
Eqs.(17), (18) and (22) that (i) $\rho _{0}(r_{NS})=0$, $p_{0}(r_{NS})=-%
\frac{\rho _{\infty }c_{\infty }^{2}}{2(16\beta ^{2}-1)}$, \ $%
c_{s}(r_{NS})=0 $ for $\frac{1}{2}<\beta <1$ (ii) $\rho _{0}(r_{NS})=0$, $\
p_{0}(r_{NS})=-\frac{\rho _{\infty }c_{\infty }^{2}}{6}$, $c_{s}(r_{NS})=%
\frac{1}{4}$ for $\beta =\frac{1}{2}$ (iii) $\rho _{0}(r_{NS})=0$, $\
c_{s}(r_{NS})=\infty $ for $\beta <\frac{1}{2}$ whereas $p_{0}(r_{NS})$
blows up only at $\beta =\frac{1}{4}$ while remaining finite elsewhere. We
notice that the behavior of fluid parameters closely resembles those at the
black hole horizon except in the minor deviation in numerical values. Only
case (iii) throws up really singular behavior. These illustrate a new
(acoustic) interpretation of a curvature singularity of general relativity.
Whether it is possible to acoustically model this singularity including the
black/wormholes in the laboratory is still anybody's guess.

We close this section by giving another particularly simple class of
solutions to Eqs. (9)-(11) given by ( $\alpha =-2$):

\begin{equation}
\phi (r)=\frac{M}{r},\psi (r)=-\frac{M}{r},\Phi =-\frac{M}{r}
\end{equation}%
where $M$ is the tensor mass as is revealed by the expansion of the metric
which coincides up to second order with the Robertson expansion [24] of a
centrally symmetric gravitational field. The solution was proposed by Yilmaz
[25] several decades ago, but it can also be obtained, like the first
example, by means of a conformal rescaling of the vacuum Brans-Dicke
equations. This is a singularity free solution representing a class of
traversable wormholes [26] with the throat occurring at $r_{th}=M>0$. The
energy density is $\rho (\Phi )=-\frac{M^{2}}{r^{4}e^{2M/r}}<0$ so that the
weak energy condition is violated. The acoustic parameters are given by

\begin{equation}
\rho _{0}(r)=\rho _{\infty },\ p_{0}(r)=0,c_{s}(r)=c_{\infty }e^{-\frac{2M}{r%
}}
\end{equation}%
This static configuration resembles a dust-like fluid. The massless limit is
trivial as $M=0$ in Eqs.(27) leads only to flat spacetime.

\section{Geodesic equations}

To what extent the formal acoustic analogue developed above makes real
sense, except satisfying some curiosity, is not clear. However, there is
always a basic but natural question, no matter whether we are concerned with
black or wormholes. We want to know if the \textquotedblleft
medium\textquotedblright\ description of gravitational interaction can be
used also to encompass other effects like, e.g., geodesic motions of general
relativity. On a hindsight, it appears that the acoustic medium formulated
in Eqs.(17), (18) [and consequently, in Eq.(22)] could be good enough for
this purpose. To describe the situation, we would digress a little, but the
arguments might still be worthwhile.

Several years ago, it was shown, in a theoretically exact formulation, that
light motion perceives gravity as an optical refractive medium [27]. In
fact, the medium notion arose as early as in the twenties of the last
century when Eddington [28] first conceived it in a somewhat approximate
manner followed by the development of what is now widely known as Gordon's
optical metric [29]. Only recently, these ideas have been revived and
applied to the theoretical investigation of acoustical/optical black holes
[30]. Along these lines of thought, and following an input from Evans and
Rosenquist [31], it has been shown that, in the spherically symmetric case,
the geodesic motion of light can be cast in an exact Newtonian $``F=ma"$
form [27] given by

\begin{equation}
\frac{d^{2}\overrightarrow{r}}{dA^{2}}=\overrightarrow{\nabla }\left( \frac{%
n^{2}}{2}\right)
\end{equation}%
in which, formally, $n(r)=c_{\infty }/c_{s}$, and $A$ is the
Evans-Rosenquist action defined by $dA=dt/n^{2}$. Note, {\it importantly},
that it is the same relation between $c_{\infty }$ and $c_{s}$ as in Eq.(18)
that determines $n$ in Eq.(29). That is, both optical and acoustical
descriptions share the {\it same }refractive index, although the two media
are governed by different sets of constitutive equations. This suggests,
{\it prima facie}, that the light ray trajectories in the optical medium
could be the same as the sound ray trajectories in the analogous acoustic
medium.

The equation (29) can be further extended to massive particle motion so that
there is now a combined geodesic equation [32]

\begin{equation}
\frac{d^{2}\overrightarrow{r}}{dA^{2}}=\overrightarrow{\nabla }\left( \frac{%
N^{2}}{2}\right) ,N=n\left[ 1-\frac{m_{0}^{2}e^{-2\phi }}{H^{2}}\right] ^{%
\frac{1}{2}}
\end{equation}%
where $m_{0}$ is the rest mass of the test particle and $H$ is the total
conserved energy that can be normalized to unity. The new refractive index $%
N $ for massive particles can be used to exactly describe all the
gravitational and cosmological kinematic effects. This reformulation of
geometry allows us to view known classical effects from an altogether
different angle. To cite an example, the anti-de Sitter geometry can be
described as Maxwell's \textquotedblleft fish-eye\textquotedblright\ lens
[33] giving rise to closed time-like curves. Moreover, Eq.(30) provides an
easy way to introduce quantum concepts in a semiclassical manner in terms of
matter de Broglie waves [32]. The method underlying the above approach has
subsequently been extended to the rotating Kerr solution by Alsing [34]. He
assumes the form of the metric on the equatorial plane ($\theta =\pi /2$) to
be of the general form

\begin{equation}
ds^{2}=-h(\overrightarrow{r})[dt-g_{i}dx^{i}]^{2}+\xi ^{-2}(\overrightarrow{r%
})\delta _{CD}dx^{C}dx^{D}
\end{equation}%
where $dl^{2}=\delta _{CD}dx^{C}dx^{D}$ is the spatial metric on the flat
two-plane. (To first order in $(m/r)$, the approximate Kerr metric can
always be written in that form with $g_{\varphi }\simeq -2ma/r$, where $a$
is the angular momentum per unit mass.) It has then been shown that the
exact geodesic equations on the equatorial slice read (no simplifying
assumptions of weak gravity or low velocity)[34]

\begin{equation}
\frac{d^{2}\overrightarrow{r}}{dA^{2}}=\overrightarrow{\nabla }\left( \frac{%
n^{2}v^{2}}{2}\right) +\frac{d\overrightarrow{r}}{dA}\times curl%
\overrightarrow{g}
\end{equation}%
where $n(\overrightarrow{r})=\xi ^{-1}h^{-\frac{1}{2}}$, $v(\overrightarrow{r%
})=n^{-1}(\overrightarrow{r})\left[ 1-\frac{h(\overrightarrow{r})}{H^{2}}%
\right] ^{\frac{1}{2}}$, $\overrightarrow{g}\equiv (g_{i})=-\frac{g_{0i}}{h}$%
 Evidently, Eq.(32) has the form of a gravitational Lorentz-like force
equation with a charge to mass ratio $(q/m_{0})$ provided we make the formal
identifications

\begin{equation}
A\equiv t%
{\acute{}}%
,A^{\mu }\equiv (\varphi ,\overrightarrow{A}),\varphi =\left( -\frac{%
n^{2}v^{2}}{2}\right) \left( \frac{q}{m_{0}}\right) ^{-1},\overrightarrow{A}=%
\overrightarrow{g}\left( \frac{q}{m_{0}}\right) ^{-1}
\end{equation}

\begin{equation}
\overrightarrow{E}=-\frac{\partial \overrightarrow{A}}{\partial t%
{\acute{}}%
}-\overrightarrow{\nabla }\varphi ,\overrightarrow{B}=\overrightarrow{\nabla
}\times \overrightarrow{A}.
\end{equation}

Complementarily, the vacuum Einstein field equations themselves can be
formally split into gravi-electric and the gravi-magnetic parts satisfying
an appropriately defined set of Maxwell's equations [35]. Returning now to
the Eqs.(30) and (32), we mention that these are useful also as an
alternative tool for dealing with other problems such as the quantum
interference of thermal neutrons in a curved spacetime [36]. As a further
implication of the medium idea, an analog of the historical Fizeau effect
can be envisaged in a gravity field [37]. These possibilities indicate that
the alternative approach captures the kinematic essence of curved spacetime
geometry in a reasonably satisfactory manner. Such analogies also provide
curious theoretical insights both into the real acoustic medium and into the
gravitational field as a result of wisdom borrowed from one regime and
employed in the other. An experimental verification of the {\it gravitational%
} Fizeau effect, although we do not know at the moment exactly how it could
be technically performed, will stand for a new test of gravity in addition
to providing indirect support to the vortex-type experiments being pursued
[8,38].

Having said all the above, we must underline a mathematical limitation of
Eqs.(30) and (32): In order to integrate these equations, an arbitrary
prescription of medium refractive index alone does not suffice; one needs to
additionally specify the metric components $g_{00}$ and/or $g_{0i}$. For
some classes of metrics, however, it might be possible to re-express them
solely in terms of the medium parameter $n$. Fortunately, in the
Schwarzschild problem which is of most practical importance, it is possible
to write $e^{-2\phi (r)}=g_{00}(m/2r)\equiv g_{00}(n)$, after solving a
cubic equation in $(m/2r)$ coming from $n(r)=(1+m/2r)^{3}/(1-m/2r)$. We can
interpret $m$ as the total mass of the spherically symmetric fluid or the
gravitating mass, as the case may be. One could then expect the geodesic
equations (30) and (32) to be generally valid for any arbitrary functional
prescription of $n$ for massless particles, and thereby $N(n)$ for massive
test particles, without the need of further specifications as to whether
this $n$ or $N$ refer to gravity field or {\it real }fluid medium. For
instance [32], the Klein-Gordon equation for spinless particles, under the
WKB approximation, leads directly to the eikonal equation $(\overrightarrow{%
\nabla }S_{0})^{2}-\overrightarrow{p}^{2}=0$, where $\left\vert
\overrightarrow{p}\right\vert =HN$, and $S_{0}(\overrightarrow{r},t)=S_{0}(%
\overrightarrow{r},t)-Ht$ is the eikonal or phase. In the acoustic regime,
from the kinematic wave front and the ray tube analyses [39], the ray
geometry satisfies the same eikonal equation \ $(\overrightarrow{\nabla }%
S_{0})^{2}-\frac{1}{c_{s}^{2}}=0,$ where we have defined $c_{s}=c_{\infty
}/N $ in harmony with the phase speed $c=c_{\infty }/n$ for massless test
particles. These arguments seem to imply that sound rays would bend in the
acoustic medium $n=n(r)$ just as much as light rays would do in the
Schwarzschild gravity. This concludes our discussion of geodesics.

\section{Summary and remarks}

Let us summarize our results. The present work is quite straightforward, but
new, with possibly more to follow in future: (1) We have developed acoustic
analogs of static traversable wormhole geometries of general relativity in
which the stress tensor is provided by the minimally coupled scalar field.
This, in turn, implies that we have essentially modelled the wormhole exotic
material and geometry together by means of usual fluid variables. (Rotating
wormholes [40] can likewise be acoustically modelled following the
developments in Ref.[6]). The wormhole analogs have been found to correspond
to energy condition satisfying fluid or ordinary matter. This result stands
in direct contrast to the fact that a very thin shell of exotic fluid is
required to wrap up the Schwarzschild black hole horizon in its acoustic
analog. The distinction appears somewhat counterintuitive since, in the
ordinary description, it is the wormhole, not the black hole, that contains
energy violating exotic matter. (2) The acoustic analog of massless
wormholes describes the singularity free Einstein-Rosen bridge model of
elementary particles in a very interesting way. In this special case, we
have found a new acoustic invariant that distinguishes the particle model
from the massive analogs. (3) The nature of curvature singularity in the
massive case has been brought forth in terms of the acoustic language. We
have seen that the acoustic behaviors at the naked singularity are not too
different from those at the horizon surface except in case (iii). Therefore,
acoustically speaking, the occurrence of a naked singularity is just as
viable as that of a horizon. This wisdom from acoustics could have
implications for Penrose's cosmic censorship conjecture in geometric general
relativity. Finally, (4) we have demonstrated that a gravitational optical
medium shares the same refractive index with the corresponding acoustic
model in the simplest case of spherical symmetry. It is argued that, in the
eikonal approximation, the equations of ray trajectories are exactly the
same both in a gravity field and in the acoustic medium so long as both are
described by the same $n=n(r)$. Therefore, the amount of bending of the rays
in two situations should be the same.

To what extent the above analog features of traversable wormholes can be
implemented in a laboratory set up is a moot question. In this context, we
know that the Hawking-Unruh radiation occurs in a fluid in transonic motion
while, in our model, the background fluid is assumed to be motionless. In
this situation, the only way to detect a horizon or a throat is to track the
phonon trajectories deep down into suitably simulated acoustic media. Then
the measurement of Fresnel transmission ($T$) and reflection ($R$)
coefficients [41], which depend only on the refractive index, could
unambiguously reveal the signatures of those distinguished surfaces present
in the medium. For the horizon, $R =1$ and $T =0$, that is, total internal
reflection, while for the throat, $R$ and $T$ have non-zero values depending
on $\beta$, satisfying $R+T=1$.

As an investigation in a different direction, it might be worthwhile to
review the problem of accretion of phantom energy ( $\rho +p<0$) into black
holes ( $\beta =1$) [42] via the mathematical machinery of standard
hydrodynamics. (Note however that a black hole's mass can evolve also due to
other reasons, for instance, by a changing Brans-Dicke scalar at the
horizon, see Ref.[43].) The phantom accretion can be viewed, within the
present framework, as a problem of superposition of two fluids: one is the
analog acoustic fluid and the other is the phantom fluid, with the accretion
ending when \textquotedblleft equilibrium\textquotedblright\ is reached. In
a similar fashion, accretion to wormholes ( $\beta >1$) [44] can be treated
using the present acoustic model. It will be of interest to see to what
extent the results from the standard geometric approach would agree with
those from the analog fluid approach. Work is underway. Notably, a recent
and very interesting work by Nojiri and Odintsov [45] shows that the phantom
stage is transient, and so is the accretion to black holes with the
consequence that the masses do not vanish.

\section*{Acknowledgments}

One of us (KKN) is indebted to Professor Liu Liao of Beijing Normal
University, China for encouragement in the present work. He also wishes to
thank Professor Ou-Yang Zhong-Can for providing hospitality and excellent
working facilities at ITP, CAS. Administrative assistance from Sun Liqun is
gratefully acknowledged. This work is supported in part by the TWAS-UNESCO
program of ICTP, Italy and the Chinese Academy of Sciences, as well as in
part by National Basic Research Program of China under Grant No.
2003CB716300. RGC would like to express his gratitude to the Physics
Department of Baylor University for hospitality. His research was supported
by Baylor University, a grant from the Chinese Academy of Sciences, a grant
from NSFC, China (No. 13325525) and a grant from the Ministry of Sciences
and Technology of China (No. TG1999075401).

\end{document}